# Temperature dependent spin dynamics in $La_{0.67}Sr_{0.33}MnO_3$/Pt bilayers.

**Biswajit Sahoo**[*], Akilan K, Katherine Matthews , Alexandre Pofelski, Alex Frano, Eric E Fullerton, Sebastien Petit-Watelot, Juan-Carlos Rojas Sanchez[*] and Sarmistha Das[*]

***Corresponding authors**

B. Sahoo, Eric E Fullerton

Center for Memory and Recording Research, University of California, San Diego, CA 92093, USA

Email: **bsahoo@ucsd.edu;**

B. Sahoo, K. Matthews, A. Frano, S. Das

Department of Physics, University of California, San Diego, CA 92093, USA

Email: **sdas@physics.ucsd.edu;**

Alexandre Pofelski:

Brookhaven National Laboratory, 98 Rochester St, Upton, NY 11973, USA

Email: apofelski@bnl.gov

Akilan K. S. Petit-Watelot, J-C. R. Sanchez

Institut Jean Lamour Université Lorraine - CNRS (UMR 7198) Campus Artem, F-54011 Nancy Cedex, France

Email: **juan-carlos.rojas-sanchez@univ-lorraine.fr**

***Complex ferromagnetic oxides such as $La_{0.67}Sr_{0.33}MnO_3$ (LSMO) offer pathways for creating energy efficient spintronic devices with new functionalities. LSMO exhibits high-temperature ferromagnetism, half metallicity, sharp resonance linewidth, low damping and a large anisotropic magnetoresistance response. Combined with Pt, a proven material with high spin-charge conversion efficiency, LSMO can be used to create robust nano-oscillators for neuromorphic computing. Ferromagnetic resonance (FMR) and device level spin-pumping FMR measurements are performed to investigate the magnetization dynamics and spin transport in $NdGaO_3$(110)/LSMO(15 nm)/Pt(0 and 5 nm) thin films ranging from 300K to 90K and compare the device performance with Py(7 nm)/Pt(5 nm) sample. The spin current pumped into Pt is quantified to determine the temperature dependent influence of interfacial interactions. The generated spin current in the micro-device is maximum at 170K for the optimally grown LSMO/Pt films. Additionally, this bilayer system exhibits low magnetic Gilbert damping (0.002), small linewidth (12 Oe) and a large spin Hall angle (≈ 3.2%) at 170K. By fine-tuning the LSMO/Pt interface quality and integrating it into the device structure, the system exhibits a fourfold enhancement in signal output for LSMO/Pt devices compared to the Pt/Py system. Such robust device level performance can pave way for energy-efficient spintronic based devices.***

Spin-based oscillators have potential for various communication applications such as radio frequency (RF) signal generator, modulator and detectors [1] [2] and can enable the development of low-power neuromorphic computing systems [3]. There are two basic classes of spin-based oscillators: spin-transfer-torque (STT) oscillators and spin-orbit-torque (SOT) oscillators [4] which differ on the mechanism of conversion of charge to spin currents. Common requisites for the development of energy efficient oscillators are high charge-to-spin current conversion, low Gilbert damping of the ferromagnetic (FM) or ferrimagnetic layer and a sharp ferromagnetic resonance (FMR) linewidth.  A smaller damping and FMR linewidth lower the critical current required for auto-oscillation, along with generation of a large output signal from the oscillator. Currently, the most commonly used materials for SOT oscillators systems are all metallic bilayers made of ferromagnet (FM)/ heavy metal (HM) (FM=CoFeB, NiFe, *etc*. ; HM=Pt, Ta, W, *etc*…) [5] [6]where the spin-orbit interactions in HM layer generates spin currents via the spin Hall effect. For FM metals, the Gilbert damping is of the order of $10^{-2}$ and FMR linewidths are of the order of 50 Oe or more [7] [8] [9] [10] [11] [12].

To enhance nano-oscillator performance, transitioning towards complex oxides has the promise for improved functionality. Half-metallic perovskite FM $La_{0.67}Sr_{0.33}MnO_3$ (LSMO) is a candidate material as the FM layer for SOT oscillators. LSMO has a very low damping [13], nearly 100% spin polarization [14], and exhibits a colossal magnetoresistance [15]. Here we report our studies of LSMO/Pt bilayers where Pt is metallic with a large charge to spin conversion efficiency. There have been reports of FMR and inverse spin Hall effect (ISHE) experiments on LSMO/Pt heterostructures showing a low damping and a reasonable charge to spin conversion [16] [17] [18] [19] [20] [21]. These experiments were conducted on unpatterned LSMO films with relatively thick layers ($t_{LSMO}$ > 20 nm), primarily at room temperature, which is near the Curie temperature ($T_C$ ) of LSMO. The key finding from these studies is that LSMO/Pt structures exhibit a low damping factor, approximately 0.007–0.01, and a spin Hall angle in the range of 1.2–2.2%. [16] [19]. Some work on the temperature dependence of spin currents on LSMO/Pt has been reported in [21],where an increase in spin current transmission with reduced temperature was observed. However, these measurements were done in LSMO(45 nm)/Pt(10 nm) on unpatterned samples of 5 x 1 x 0.5 mm, which can downplay the contribution of interfacial effects. To date, the temperature dependence of spin-charge conversion in LSMO/Pt thin films below 300 K with $t_{LSMO}$<20 nm, as detected by spin pumping voltage at the device level, remains unexplored.

For many spintronic and neuromorphic applications, one requires robust nano/micron size devices formed from low thickness of the FM and HM layer to enhance interfacial effects [22]. The properties of thin film system may be altered after the harsh conditions of device fabrication such as exposure to high vacuum environments for contact deposition, high energy ions bombardment for etching and sonication in chemicals and solvents. For perovskites, such environments can cause generation of defects by various pathways such as sample damage [23], oxygen concentration change [24] *etc*., which may give sub-optimal device-based performance, despite showing robust properties in unpatterned films.

The increased magnetization and lower damping [25] [26] of LSMO thin films at lower temperatures can potentially pump larger spin currents into the adjacent HM, enabling larger output voltage signals and energy efficient device performance. Thus, it is worthwhile investigating the temperature dependent magnetization dynamics of LSMO/ Pt system and the role of the interface for spin transport, both at the thin film and the device level. The inclusion of Pt will lower the device resistance enabling robust device stability and optimal response to currents. The samples considered here are LSMO(15 nm) films and LSMO(15 nm)/Pt(5 nm) bilayers grown on $NdGaO_3$(*110*)(NGO(*110*)) single-crystal substrates. NGO(*110*) substrates were used as they do not have the high relative permittivity and dielectric loss of some commonly used oxide substrates, such as $SrTiO_3$, which can otherwise cause significant current shunting in substrates at high frequency even at room temperature, and may strongly affect spin-torque measurement resulting in an inaccurate estimation of charge-to-spin conversion efficiency [27].

We perform temperature dependent FMR on unpatterned films where we detect the absorbed microwave power, and spin-pumping FMR (SP-FMR) on patterned microdevices where we detect the spin pumping voltage. We quantify the evolution of the linewidth, effective magnetization, intrinsic Gilbert damping, spin current density and the spin Hall angle (SHA) versus temperature to find the best conditions to create an energy efficient spin Hall nano-oscillator and related devices. We further provide comparison between LSMO/Pt and the conventionally used Py/Pt devices fabricated under same conditions to show the efficacy of the LSMO/Pt system in device-based applications. Our results also provide concrete insights on the role played by the interfacial layer for spin transport across the LSMO/Pt interface.

Epitaxial LSMO films were grown using pulsed laser deposition on NGO(*110*) substrates. Figure 1 shows the transmission electron microscopy (TEM) image of the NGO/LSMO sample displaying a highly epitaxial cube -on -cube growth of LSMO on NGO. LSMO grows in the *001* direction as demonstrated by the X-ray diffraction spectra (Fig. S2(a)) and the electron diffraction (Fig. 1 (b)).

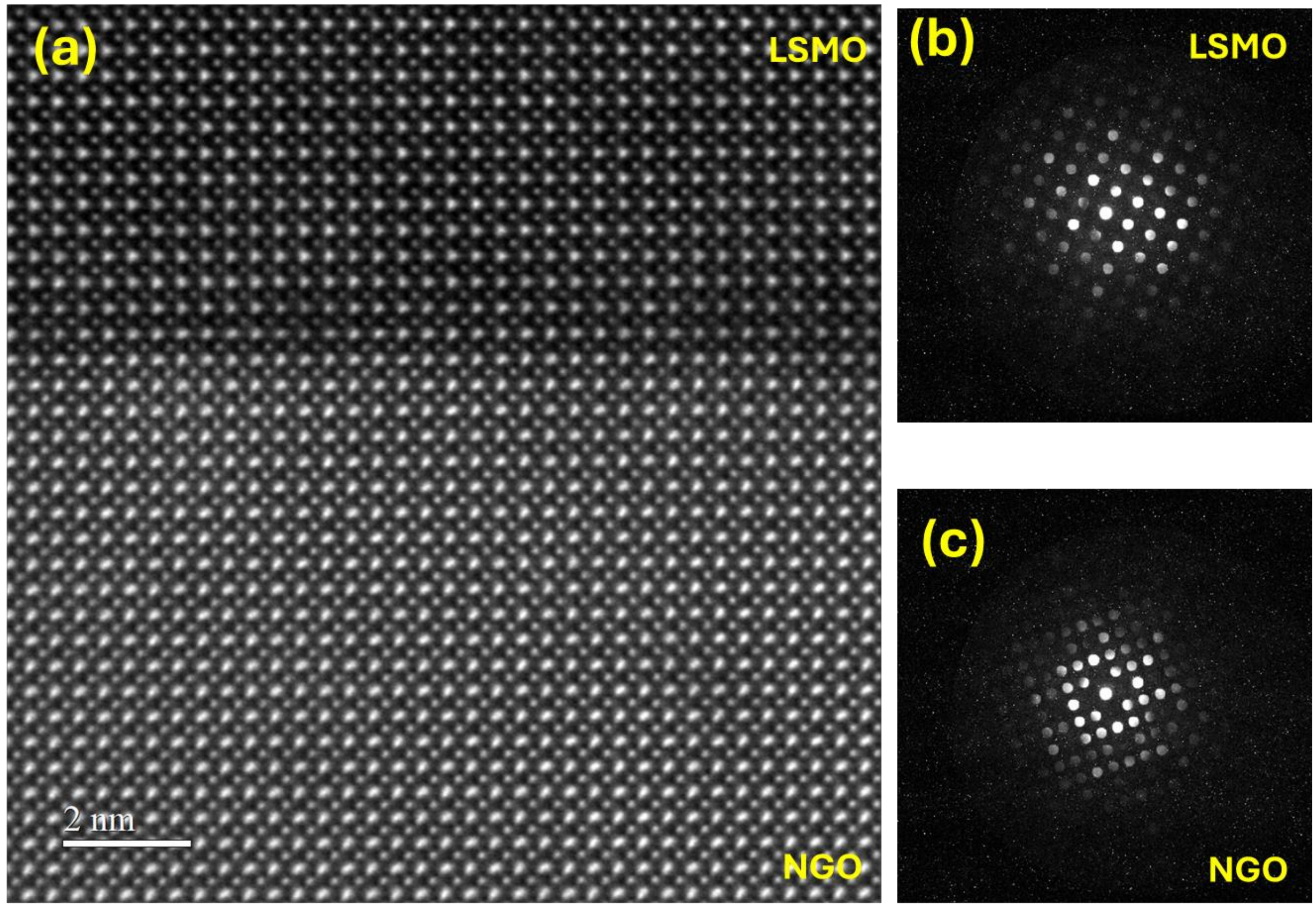

*Figure 1 TEM image of NGO(110)/LSMO(15) showing epitaxial growth. (b) and (c) show the electron diffraction pattern for LSMO and NGO respectively establishing a highly crystalline structure.*

The sample was transported quickly through the air into a sputtering chamber where Pt was deposited. The sample fabrication details are provided in the supplementary.

The resistance-vs.-temperature measurements (Figs. S4(a)) for LSMO show a typical semi-metallic behavior, with resistivity values 3mΩ-cm at room temperature (RT) similar to those reported in the literature [28]. For LSMO/Pt (Fig.S2(b)), the resistivity is 28μΩ.cm at RT, is similar to that measured for Pt thin films [10] [29], indicating that the majority of the current shunts through the Pt layer.

The FMR measurements were performed using a field-modulated, flip-chip co-planar waveguide, in conjunction with a cryogenic set-up. FMR measurements were performed at a frequency range of 5-15 GHz at 1 GHz intervals at temperatures ranging from 300K to 90K. Various magnetic parameters were obtained at all temperatures from analysis of the FMR spectra. Figure. 2 shows the FMR spectra of LSMO and LSMO/Pt at 5GHz for different temperatures. At 300K, the signal consists of two modes which merge into a single mode at lower temperatures (T<=250K). This may be due to the presence of multiple collective oscillation modes as a result of weaker exchange interaction of LSMO at RT, which increases at lower temperatures [30] enabling the emergence of a single sharp peak. Increase in the saturation magnetization ($M_s$) shifts the resonance field to lower values with reducing temperature. A similar trend is observed in the LSMO/Pt sample. For fitting of a single FMR mode, we use the sum of symmetric and anti-symmetric Lorentzian function:

$$\frac{dP}{dH} = off + mH + S_1 \frac{\Delta H^2 - 4(H - H_{res})^2}{(4(H - H_{res})^2 + \Delta H^2)^2} - 4A_1 \frac{\Delta H(H - H_{res})}{(4(H - H_{res})^2 + \Delta H^2)^2} \quad (1)$$

where $H_{res}$ is the resonant field, $\Delta H$ is the full width at half maximum; $m$ and $off$ are the slope and offset, respectively. For the 300-K spectra, which has two peaks, we employ a two-mode fitting approach using two pairs of symmetric and anti-symmetric Lorentzian functions. The results of the fits are shown in Figs. 2 and S5.

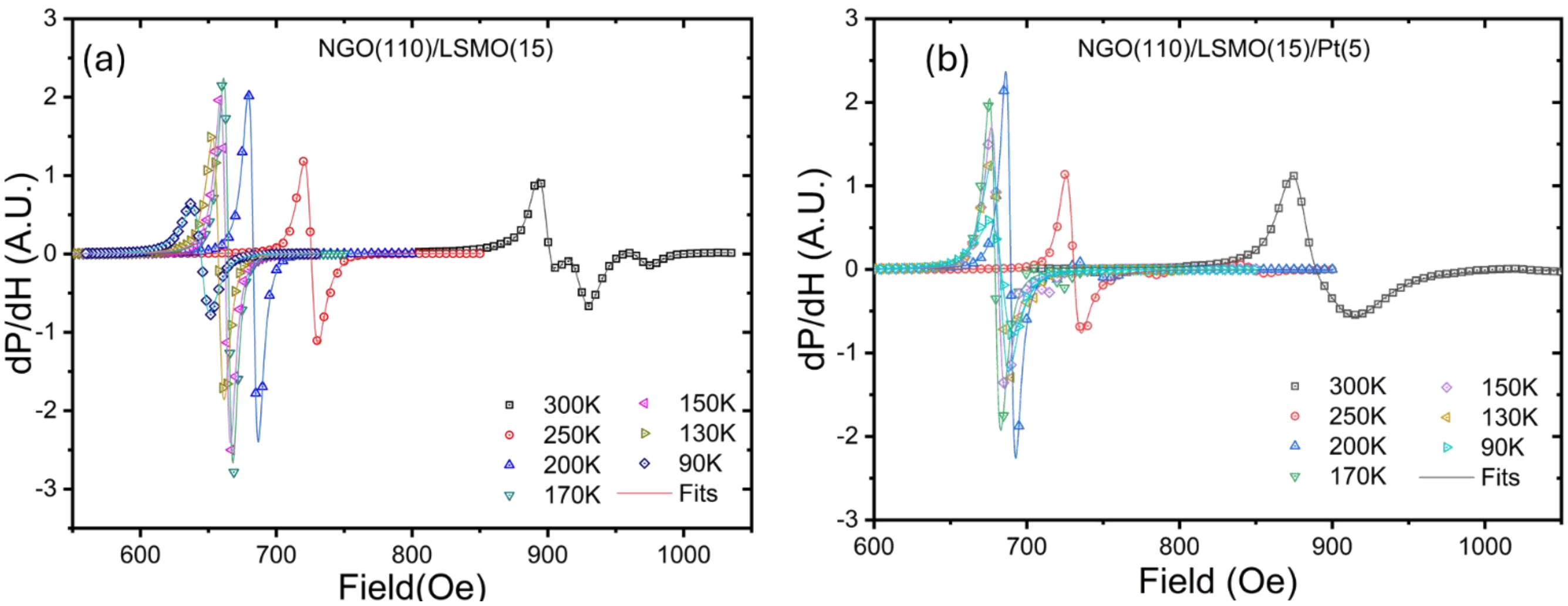

*Figure 2 (a) and (b) show the temperature dependent FMR spectra measured at 5 GHz for LSMO and LSMO/Pt, respectively. Solid lines are the fits to the data. The lineshape for both the samples shows a similar variation with multiple peaks at 300K, which combine into a single mode at lower temperatures.*

We fit the $H_{res}$ vs. frequency data to the Kittel equation [31]:

$$f = \frac{\gamma}{2\pi} \sqrt{(H_k + H_{res})(H_K + H_{res} + 4\pi M_{eff})} \quad (2)$$

where $\gamma$ is the gyromagnetic ratio, $H_k$ is the in-plane uniaxial anisotropy field and $M_{eff}$ is the effective magnetization. From the fits (Figs. S5(a) and (b)) we see that the resonant fields for the LSMO film are approximately the same as that of LSMO/Pt indicating the Pt layer is not significantly affecting the underlying LSMO. Additionally, the similar $M_s$ values for LSMO and LSMO/Pt (Fig. S7 (a)) indicate a negligible dead layer at the LSMO/Pt interface. The $M_{eff}$ obtained from FMR is lower with respect to the $M_s$ obtained from VSM experiments (Fig. S7 (b)). The lower $M_{eff}$ value indicates an additional perpendicular magnetic anisotropy (PMA) as $M_{eff}= M_s-H_a/4\pi$ , where $H_a$ is the out-of-plane (OOP) anisotropy field that is present in both the LSMO and LSMO/Pt films. This is confirmed by OOP magnetometry of LSMO, where we observe increasing contribution of $H_a$ with reduced temperatures (Fig. S7 (d)) as the OOP saturation field is much lower than $4\pi M_s$. Such PMA has been attributed to magnetostriction due to interfacial strain [26] [27]. Given that $M_{eff}$ is very similar to LSMO and LSMO/Pt structures, this suggests there are no significant contributions of interfacial anisotropy from the LSMO/Pt interface.

To obtain the intrinsic damping, we fit the line width vs frequency to the following equation:

$$\Delta H = \Delta H_0 + \frac{4\pi\alpha}{\gamma} f \tag{3}$$

where $\Delta H_0$ is the inhomogeneous linewidth broadening, $\alpha$ in the intrinsic Gilbert damping, and $\gamma$ is the gyromagnetic ratio obtained from Eq. (2). Figs. 3(a) and (b) show the fits to the $f$ vs $\Delta H$ curves using Eq. (3). It should be noted that there can be additional non-linear contributions to the linewidth broadening [32] [33]. However, in our case, both LSMO and LSMO/Pt show a linear behavior of $\Delta H$ with respect to frequency and thus these effects were not included in the analysis.

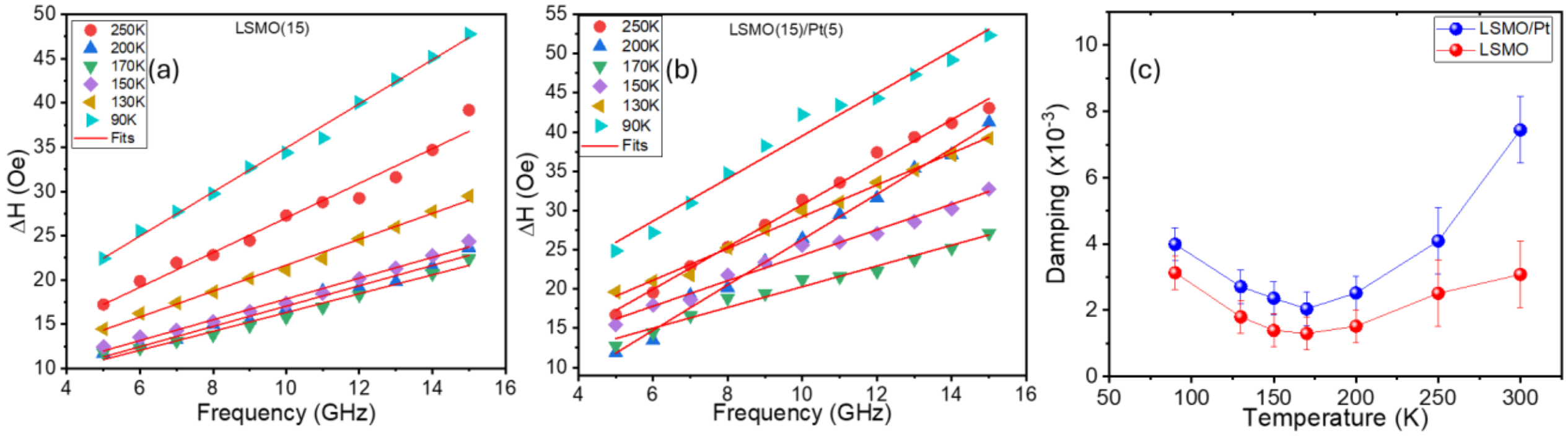

*Figures 3 FMR results. (a) and (b) show the linewidth vs frequency plots for LSMO and LSMO/Pt, respectively (points) and the solid lines are fits to Eq. 3. The fits show a linear behavior indicating minimal contribution of two magnon scattering and other non-linear processes. (c) shows the obtained Gilbert damping variation with temperature.*

The variation of the Gilbert damping magnetic with respect to temperature for LSMO and LSMO/Pt is shown in Fig. 3(c). For LSMO, the damping is 0.003 at room temperature, which is comparable to values reported in the literature for similar LSMO thickness values [25] [20] [19]. The damping decreases with decreasing temperature, reaching a minimum value of 0.0013 at 170K and then increases to 0.0031 at 90K. Observation of such minima in the damping has also been reported in LSMO thin films [25] [26]. LSMO/Pt follows a similar trend, albeit with an increased value of α (0.007 at 300K, 0.002 at 170K and 0.004 at 90K), which can be attributed to additional spin pumping into the Pt layer [7] [20]. In contrast, the damping for a reference Py(7)/Pt(5) bilayer over the same temperature range remains nearly constant at 0.015 (Fig. S8(a)). Additionally, the inhomogeneous broadening is low in LSMO and LSMO/Pt indicating a high-quality magnetic layer growth in both samples. (Fig. S6(b)).

The minima in damping with temperature can be explained by the Kambersky's torque correlation model [34] and its extensions [35] [36], which incorporate contribution of conductivity-like (based on the model of propagating spin waves in a breathing Fermi surface) and resistivity-like (based on spin flip due to magnon absorption) components into the intrinsic Gilbert damping. Fits to this model are shown in Fig. S6, which enables us to gauge the contribution of the above-mentioned effects to the intrinsic damping.

The increase in damping at lower temperatures may also have a contribution from the coupling of the magnetic moments of the metallic part of LSMO to a phase separated magnetically active (PSMA) region [25]. The PSMA region refers to the portion that departs from the metallic FM property of LSMO and is claimed to have uncompensated AFM spins [25]. Since the bulk of the film is supposed to be a metallic FM, the PSMA layer may be primarily located at the top and bottom interfaces of LSMO film [37] [25]. From EELS spectra, we confirm a different phase existing at the bottom of the LSMO layer, which is due to interdiffusion of Nd into LSMO (Fig. S3). This layer has a shorter spin relaxation time and can act as a spin sink, resulting in an increase in the damping [25] [26]. Kambersky's model cannot account for the interactions arising from this layer which results in deviation from the damping fitting by this model (Fig. S6(a)). The role of PSMA layer for spin scattering has been indirectly confirmed by taking into account the deviation from fitting to Kambersky's model in [25] [26] and is more prominent for thinner LSMO ($t<5$nm). In the later sections, we provide direct evidence of the role of this layer in the transmission of spin current into Pt.

For quantifying the spin-to-charge conversion efficiency and the effects of the PSMA layer on spin transmission, the LSMO/Pt films were patterned into devices via a 3 step optical lithography process for spin-pumping FMR (SP-FMR) measurements [38] [39]. These measurements were performed from 5 to 15 GHz at 1 GHz intervals and 31.6 mW (15dBm) of input power, at the same temperatures as the FMR measurements. Figure S1 shows the schematic of the set-up. An RF current is applied through ground-signal-ground (G-S-G) probes onto a waveguide patterned directly on the sample, which includes a 50-Ω terminator for impedance matching. Simultaneously, the external in-plane magnetic field applied is swept perpendicular to the direction of the radio frequency (RF) wave oscillation. The magnetic moments in the FM precess due to the resonance due to RF and DC field, and pure spin current is pumped into the adjacent HM. This spin current is converted into charge current by the inverse spin Hall effect in the HM layer, resulting in a measurable voltage that is detected by gold probes. The resulting lineshape consists of symmetric and antisymmetric Lorentzian components, with a voltage maximum (or minimum) at the resonance field. The signals are fit to the following from:

$$V_{total} = V_s \frac{\left(\frac{\Delta H}{2}\right)^2}{\left(\frac{\Delta H}{2}\right)^2 + (H - H_{res})^2} + V_A \frac{\left(\frac{\Delta H}{2}\right)(H - H_{res})}{\left(\frac{\Delta H}{2}\right)^2 + (H - H_{res})^2} \ . \quad (4)$$

Here $V_s$ is the symmetric contribution of the voltage which has major contributions from spin-pumping and anisotropic magnetoresistance (AMR). $V_A$ is the anti-symmetric part, which consists of effects from AMR, anomalous Hall and planar Hall effects (AHE,PHE) [40].

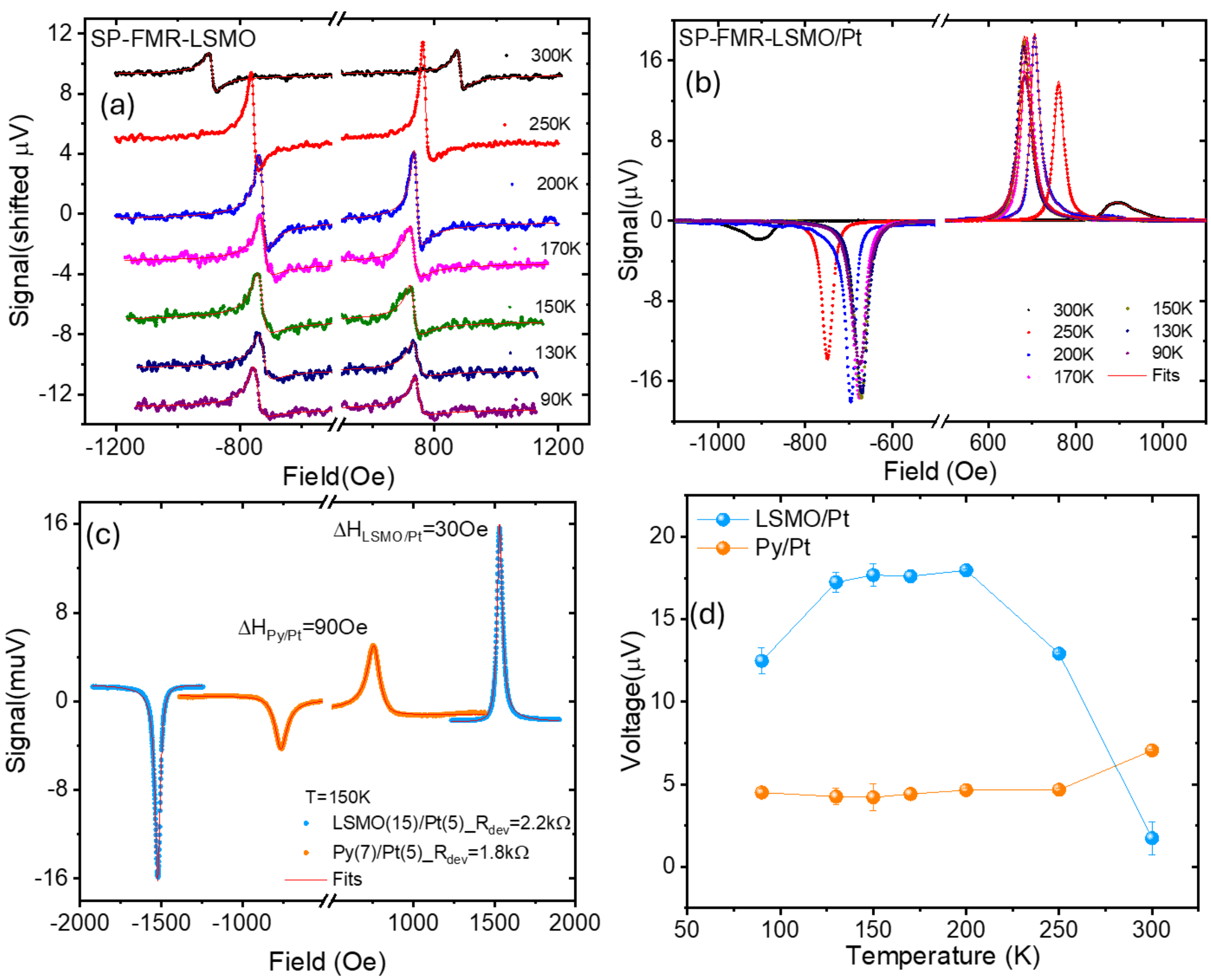

*Figure 4 SP-FMR results (a)-(b) show the SP-FMR spectra at 5 GHz for LSMO and LSMO/Pt respectively. The signal for LSMO has been shifted for clarity. For LSMO/Pt, the signal increases with lower temperatures and is dominated by spin pumping contribution. The red solid lines indicate fits to Eq 4. (c) shows comparative SP-FMR signal strength between Py(7)/Pt(5) and LSMO(15)/Pt(5) on NGO(110) substrate at 150K and 8 GHz (d) shows the comparison of the SP-FMR voltage for LSMO/Pt and Py/Pt at 8GHz for different temperatures.*

We first perform SP-FMR measurements at 5 GHz and 31.6 mW RF power at different temperatures on patterned LSMO films (Fig. 4(a)) to rule out any contribution to the spin pumping signal arising from single layer LSMO. First, we focus on the signal strength. The overall peak-to-peak voltage amplitude is less at 300K (<1uV) and attains a maximum around 200K (3uV), which can be attributed to the increase in $M_s$ and a stronger exchange interaction strength at lower temperature. Below 170 K, the signal strength decreases. This behavior suggests a larger dissipation of the angular momentum below 170 K. This can be within the bulk of the film and/or due to increased interaction with the PSMA layers. We can see that at all temperatures, the signal has the same shape in the positive and the negative field directions, indicating negligible spin-pumping contribution. This signal mainly arises from AMR effect, which has symmetric and antisymmetric contributions, and the AHE effect which has a dominant anti-symmetric contribution in this measurement configuration [40] [12]. As we go down in temperature, the anti-symmetric contribution reduces, indicating a reduction in the AHE effects. This can be due to the combined effect of reduction of resistivity of LSMO [41] with decreasing temperature and increased absorption of the angular momentum within the bulk/PSMA layers. These measurements confirm the increased dissipation of spin angular momentum at lower temperatures, and that the PSMA layer does not convert the scattered spins into a measurable spin-pumping voltage.

Figure 4(b) shows the SP-FMR signal of LSMO/Pt. A stark contrast is observed in the lineshape and signal strength. The signal sign exhibits an odd symmetry with respect to positive and negative field directions and has a dominant symmetric component. The symmetric component is 5-10 times greater in signal strength with respect to LSMO. This indicates that LSMO/Pt total voltage has a significant contribution from spin pumping [42] [40]. Moreover, the observed behavior is primarily governed by the properties at the LSMO/Pt interface.

For 300K, the LSMO/Pt signal is quite broad and is composed of two modes, similar to that observed in the FMR spectrum. This may be due to multiple non-uniform modes due to a lower exchange interaction energy and/or magnetic inhomogeneities present in LSMO [18]. However, as the temperature decreases, the signal values increase drastically from 2μV at 300K to 18μV at 200K with presence of a sharper single mode. This behavior is similar to the FMR spectrum, which can be attributed to a stronger FM exchange coupling in LSMO at lower temperature, enabling uniform collective oscillation, and consequently a higher magnitude of spin pumping into Pt. The signal attains a maximum value between 200-170K, which then reduces to 12 μV at 90K. The linewidths from SP-FMR and FMR show a similar trend to the damping. The SP-FMR (FMR) linewidth reduces drastically from 64 Oe (62Oe) at 300K to a minimum of 24 Oe (11Oe) at 200K and then increases to 38 Oe (25Oe) at 90K. The increased linewidth of LSMO/Pt in the SP-FMR measurements with respect to FMR spectra may be due to some sample differences during device fabrication. The linewidth variation of single layer LSMO devices shows more randomness as opposed to LSMO/Pt devices (Fig(S6.(c)). This indicates that deposition of the Pt layer (in spite of being *ex situ*) protects the device from extensive sample damage, which is advantageous for device fabrication. Further, upon comparing the relative signal strength and device resistance of LSMO(15)/Pt(5) with that of Py(7)/Pt(5), we clearly see the advantages of LSMO/Pt system (Fig. 4(c)). The signal strength at 150K is 4 times higher in LSMO/Pt (17.8μV) as opposed to Py/Pt (4 μV) for comparable device resistances. It is also to be noted that the linewidths and damping for Py(7) and Pt(7)/Py(5) system are consistently higher than LSMO and LSMO/Pt respectively and remains almost constant with temperature (Fig. S8(b)). This shows the merit of using LSMO/Pt in device-based applications.

The spin Hall angle ($\theta_{SHA}$) is the figure of merit to determine the charge to spin conversion ratio. We obtain $\theta_{SHA}$ through the following equation [43]:

$$I_{SP} = \theta_{SHA}\, J_s w \lambda_{sf} \tanh\left(\frac{t_{pt}}{2\lambda_{sf}}\right). \tag{5}$$

Here $I_{SP}$ is the spin-pumping current obtained by $(V_{S+}\text{-}V_{S-})/(2R_{dev})$. $V_{S+(-)}$ is the spin pumping voltage at positive (negative) fields. $R_{dev}$ is the resistance of the device, $t_{Pt}$ is the thickness of Pt (5 nm), $w$ is the width of the device (10 μm) and $\lambda_{SF}$ is the spin diffusion length of Pt, which we have assumed to be 3.4 nm [7]. $J_s$ is the spin current density is determined by the following relation [43]:

$$J_s \approx \left(\frac{g_{eff}^{\uparrow\downarrow}\hbar}{8\pi}\right)\left(\frac{h_{rf}\gamma}{\alpha}\right)^2 \left[\frac{4\pi M_{eff}\gamma + \sqrt{\left(4\pi M_{eff}\gamma\right)^2 + (4\pi f)^2}}{\left(4\pi M_{eff}\gamma\right)^2 + (4\pi f)^2}\right]\left(\frac{2e}{\hbar}\right). \tag{6}$$

Here $e$ is the electronic charge and $h_{RF}$ is the RF field experienced by the sample which was calibrated beforehand for all RF frequencies [44] and $g_{eff}$ is the effective spin mixing conductance given as $g_{eff}^{\uparrow\downarrow} = \frac{4\pi M_s \mathrm{d_{FM}} \Delta\alpha}{g\mu_B}$ [45], where $g$ is the g-factor obtained from the Kittel fittings of sample LSMO/Pt (Fig. S5(c)), $\mu_B$ is the Bohr magneton. $\Delta\alpha$ is the difference between the damping of samples LSMO/Pt and LSMO ($\alpha_{LSMO/Pt} - \alpha_{LSMO}$) obtained by FMR, $d_{FM}$ is the thickness of the ferromagnetic layer (15.0 nm).

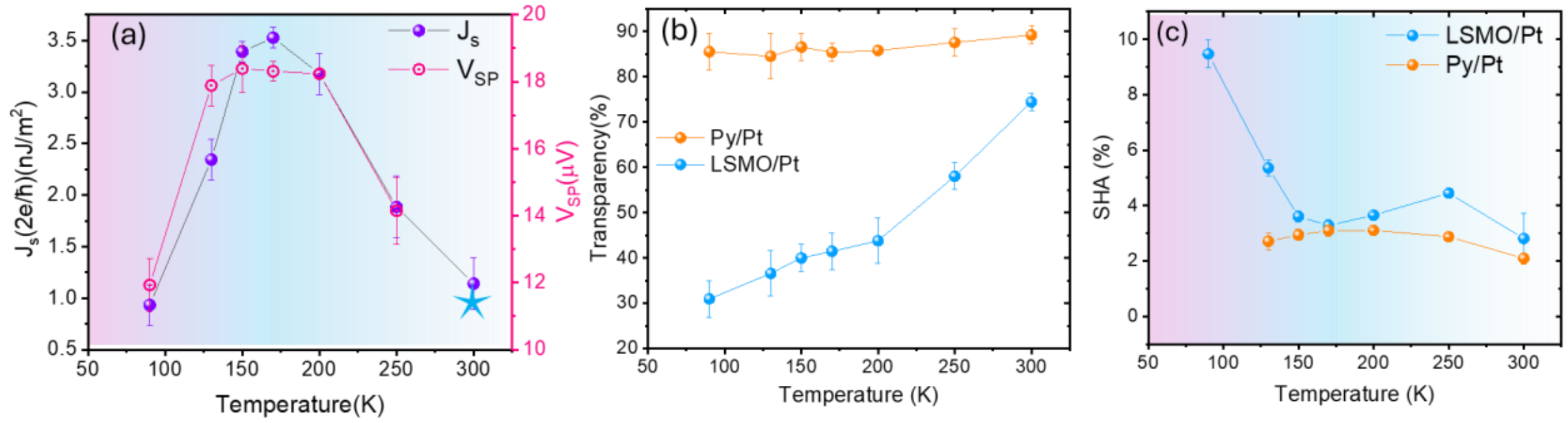

*Figure 5 SHA determination from SP-FMR results. (a) shows the variation of the spin current density into the Pt layer with temperature for LSMO/Pt along with the spin pumping voltages at different temperatures at 5GHz. $V_{SP}$ for 300K is not plotted as the two-mode fitting may not provide a reliable value unlike single mode fitting for lower temperatures. The star indicates a value of 0.75 nJ/m² for LSMO/Pt system as reported in ref [19].(b) Shows the variation of the interfacial transparency for LSMO/Pt (c) shows the variation of the spin Hall angle with respect to temperature.*

The variation of $J_s$ vs. temperature for LSMO/Pt is shown in Fig. 5(a). We see that the spin current increases with decreasing temperature down to 170K, which can be attributed to the increased magnetization of the LSMO, resulting in greater spin pumping into the system. Below 150K, there is a significant interaction between the metallic LSMO and the PSMA layer at LSMO/Pt interface, resulting in absorption of some of the spin current being pumped into Pt, which is also reflected in the variation of the $V_{SP.}$ To further shed light on this behavior of $J_s$, we track the behavior of the spin transparency ($T$) at the LSMO/Pt interface which is given as [46], [10]:

$$T = \frac{g_r \, tanh\left(\frac{t_{Pt}}{\lambda_{Pt}}\right)}{g_r \, coth\left(\frac{t_{Pt}}{\lambda_{Pt}}\right) + \frac{h}{2e^2 \, \rho_{Pt}\lambda_{Pt}}} \tag{7}$$

Here $g_r$ is the real part of $g_{eff}$ and we have assumed $g_r = g_{eff}$ [10]. The interfacial transparency decreases with temperature from 76% at RT to 30% at 90K at the LSMO/Pt (Fig 5(b)), which can be interpreted as a direct consequence of the PSMA layer scattering the spin currents going into Pt. However, the simultaneous increase of the $M_s$ and FM exchange interaction strength leads to stronger moment precession of LSMO, thus dominating the reduction in the spin transparency. However, the $M_s$ saturates below T< 150K and the scattering of the spin current becomes dominant leading to a decrease in the $J_s$. This is in stark contrast to the Py/Pt system, where the spin transparency remains almost constant throughout the temperature range.

Figure 5(c) shows the variation of the spin Hall angle with respect to temperature. At 300 K, we obtain a $\theta_{SHA}$ of ≈ 2.3%. It then increases and varies from 4.4 to 3 % from 250K-150K. The $\theta_{SHA}$ then increases to

≈ 9% at 90K. The large error margin of SHA at 300 K is due to low spin pumping signal, the strength of which is split between two modes. At lower temperatures, only one mode emerges with a strong spin pumping voltage, resulting in a more consistent value of the $\theta_{SHA}$ (down to T=150 K). However, upon further decreasing the temperature (130-90 K), the interaction between the metallic part of LSMO with the PSMA region increases, which further enhances the damping. From Eq. 6, we can see that the $\theta_{SHA}$ is directly proportional to the overall damping of the system. Thus, this additional increase in damping leads to a seemingly increased value of $\theta_{SHA}$ which may not reflect the true spin-charge conversion efficiency. In the temperature regime where this interaction is inconsequential (250K<T<170K), the obtained value of the spin Hall angle between is consistent with the measured Py/Pt sample and also with many reports involving Pt/FM systems [7] [47] [48] [38] [10].

Despite the *ex-situ* nature of fabrication, we focused on carefully optimizing the thickness and quality of the LSMO layer and found that the LSMO/Pt exhibit robust SP-FMR signals at all temperatures ranging from 300K to 90K. We report the lowest observed damping for LSMO(15)/Pt(5) system at 170K (0.0020 ± 0.0005), almost 2-3 times lower than values reported to date [17] [20] [16] [18], for systems with comparable or larger LSMO thicknesses. The thin LSMO and Pt layers allow for the observation of pronounced interfacial effects in the system, leading to generation of strong spin pumping voltage signal in micron size devices. Further, we report a sharp FMR linewidth at 5GHz (12 ± 1 Oe at 170K) with a robust spin Hall angle (3.2 ± 0.4 %). We observe a PSMA layer form at the NGO/LSMO interface via EELS measurements. While the PSMA layer may contribute to an increase in damping, it does not contribute to spin pumping through the entire temperature range. We further provide concrete evidence of the role of PSMA layers on the spin current transmission into Pt as a function of temperature. The spin current pumped into Pt is the maximum at 170 K (3.5 ± 0.1 nJ/$m^2$) which then decreases with reducing temperature due to spin scattering effect arising from the increased interaction between spins of metallic LSMO with that of top PSMA layers of LSMO. This is corroborated by the interfacial transparency calculations and comparison with Py/Pt system. Comparison of device performance of LSMO/Pt with popularly used Py/Pt for 150K<T <200K show a 4 times larger signal output with 3 times lower linewidth for similar device resistances. Further the thin film properties show an 8-fold reduction in damping for LSMO/Pt as opposed to Py/Pt. All these properties point to the robustness of using this bilayer system as an efficient spintronic device for applications such as low-power nano-oscillators.

**Acknowledgements:**

This work was primarily supported as part of Quantum Materials for Energy Efficient Neuromorphic Computing (Q-MEEN-C), an Energy Frontier Research Center funded by the U.S. Department of Energy (DOE), Office of Science, Basic Energy Sciences (BES), under Award # DE-SC0019273.

This material is based upon research partially supported by the Chateaubriand Fellowship of the Office for Science & Technology of the Embassy of France in the United States.

Devices were patterned at Jean Lamour Institute's cleanroom platform, MiNaLor, which is partially funded by the Grand Est region via the project RANGE. We also acknowledge the support from EU-H2020-RISE project Ultra Thin Magneto Thermal Sensoring ULTIMATE-I (Grant ID. 101007825).

This work was partially supported by the project "Lorraine Université d'Excellence" reference ANR-15-IDEX-04-LUE, through the France 2030 government grants EMCOM (ANR-22-PEEL-0009), and PEPR SPIN – SPINMAT ANR-22-EXSP-0007.

# Supplmentary Information for Temperature dependent spin dynamics in $La_{0.67}Sr_{0.33}MO_3$/Pt bilayer.

**Biswajit Sahoo**[1,2*], Akilan K[3], Katherine Matthews[2], Alex Frano[2], Eric E Fullerton[1], Sebastien Petit-Watelot[3], Juan-Carlos Rojas Sanchez[3*] and Sarmistha Das[2*]

15-nm-thick LSMO films were grown on $NdGaO_3$(110) (NGO) substrates by pulsed laser deposition. The substrate temperature during deposition was kept constant at $950^0$C. The films were grown at an $O_2$ pressure of 200 mTorr and annealed at the same temperature for 15 minutes at 7.6 Torr $O_2$ pressure and then cooled down to room temperature at the same pressure. Freshly grown, pristine LSMO films served as the base for subsequent *ex-situ* deposition of a 5-nm Pt layer to create desired bilayer structures. Pt was sputter deposited at room temperature, at an Ar deposition pressure of 2.7 mTorr and with a base pressure better than 7 x $10^{-8}$ Torr. NGO(*110*)/Py(7)/Pt(5) was sputtered in situ in the same conditions mentioned above.

For Fabricating SP-FMR devices, the film was patterned into devices of 500 x 10 µm wires via Ar ion etching. Next, 200nm of $SiO_2$ was deposited on the entire substrate, leaving areas for gold contact deposition. Finally, 5-nm Ti and 150-nm Au layers were deposited by evaporation to fabricate ground-signal-ground waveguides and the contacts by lift-off. For measurements, the devices were placed in a cryo-chamber with 4 probes: two for sending RF currents and two for measuring the resulting voltage. An external field perpendicular to the RF current was swept, and the resulting voltage (due to spin rectification effects) was recorded by a nano-voltmeter. The schematic of the setup is shown in Figures S1

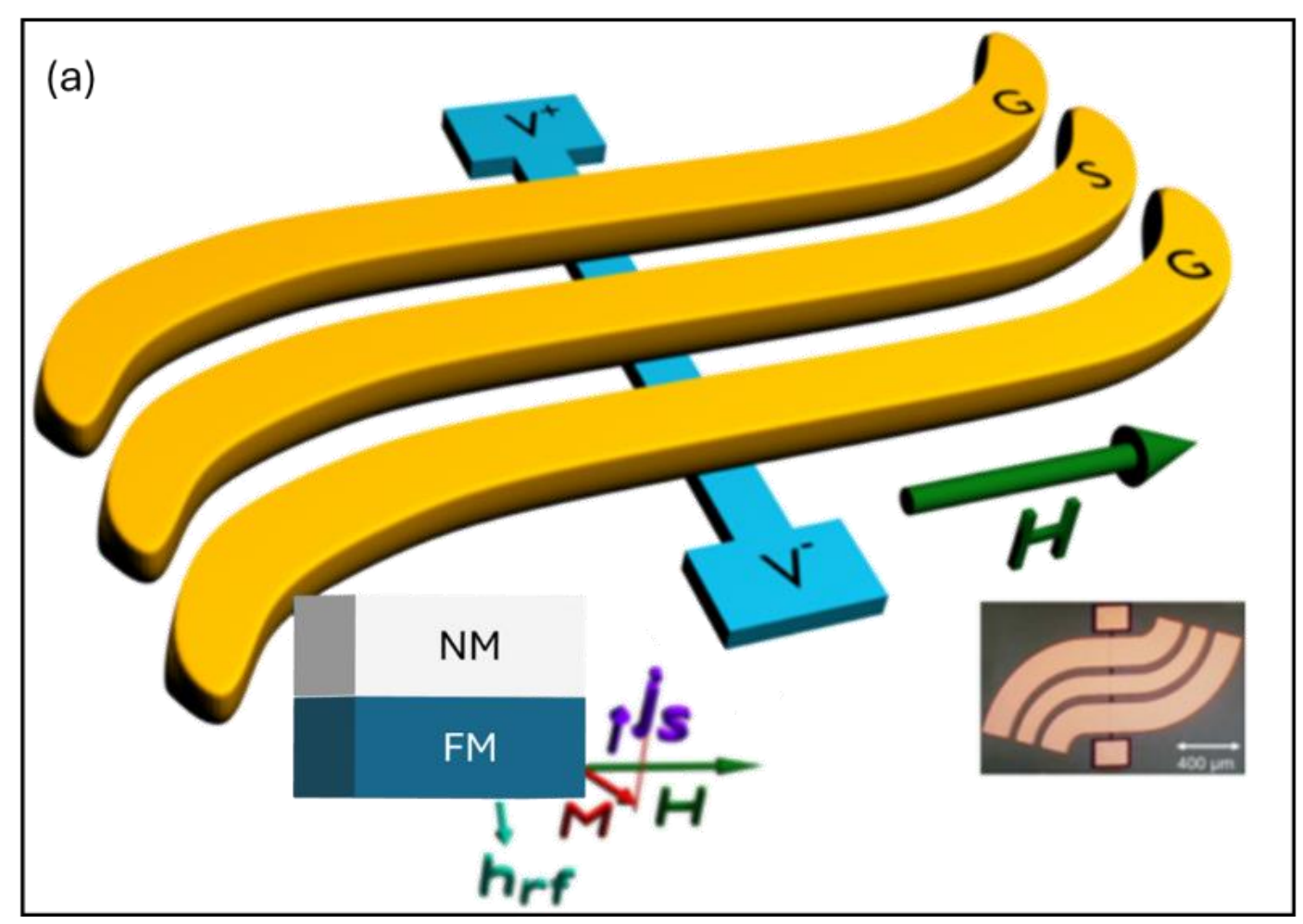

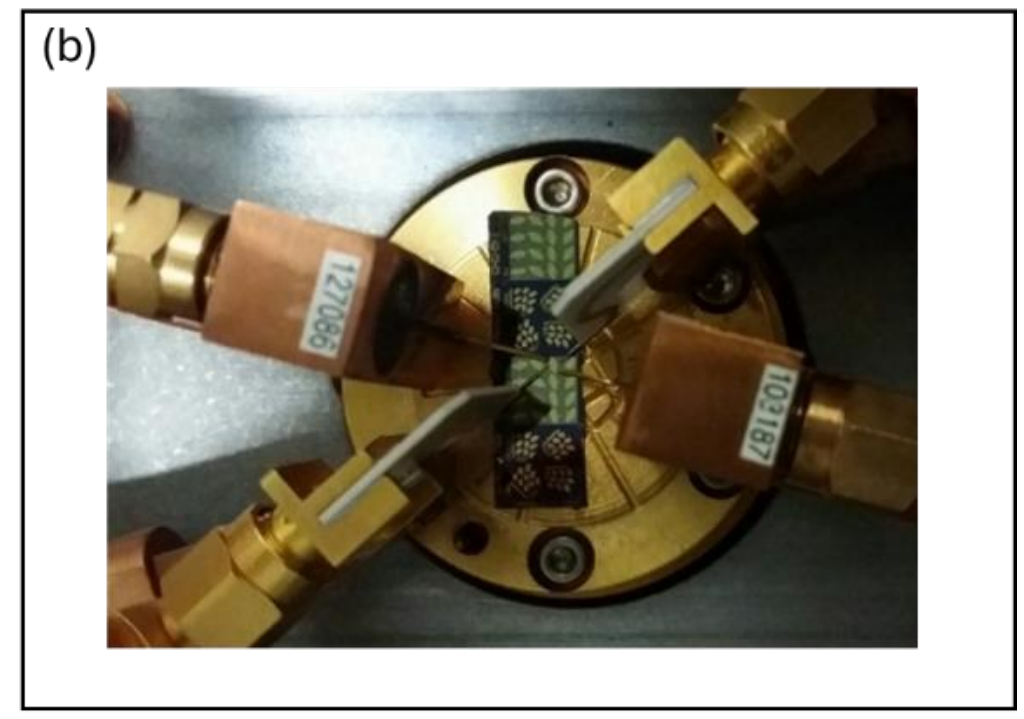

*Figure S2 (a) shows the schematic of the SP-FMR measurement set-up. Inset bottom left is a micrograph of the fabricated device. (b) shows the probe placement for measurement of the SP-FMR signals. The bright yellow blocks (bottom left and top left) are the RF probe and the 50Ω terminator probe respectively. The top left and bottom right are the voltage probes.*

The crystallinity of LSMO was determined via X-Ray Diffraction (XRD) using Rigaku Smartlab systems which employs Cu-Kα X-rays of wavelength 1.54 Å. S1 shows the full XRD scan with a rocking curve about the 002 LSMO peak. The FWHM from the rocking curve is 0.0954 revealing a highly epitaxial film growth.

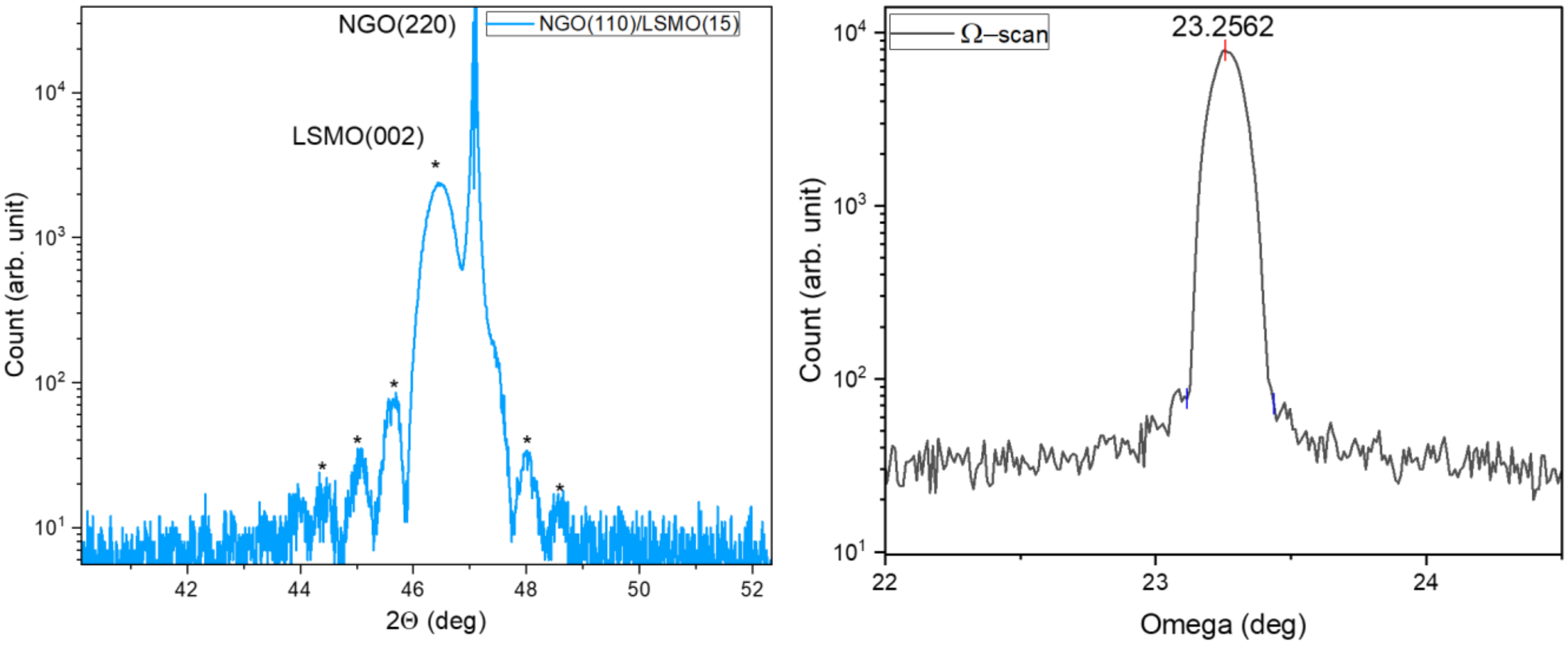

*Figure S3 (a) shows the XRD spectra of the samples with the LSMO(002) peak at ≈ 46.5$^0$.The * mark the Laue fringes associated with LSMO. (b) shows the rocking curve around LSMO(002) peak with FWHM of 0.094$^0$.*

The electron energy loss spectroscopy (EELS) measurements show a detailed interfacial environment at the NGO/LSMO interface (Fig. S3). The bottom interface of LSMO is darker in color indicating a possible different interfacial phase. The EELS spectra confirm the diffusion of Nd into LSMO at the lower interface creating a different phase from the rest of the film.

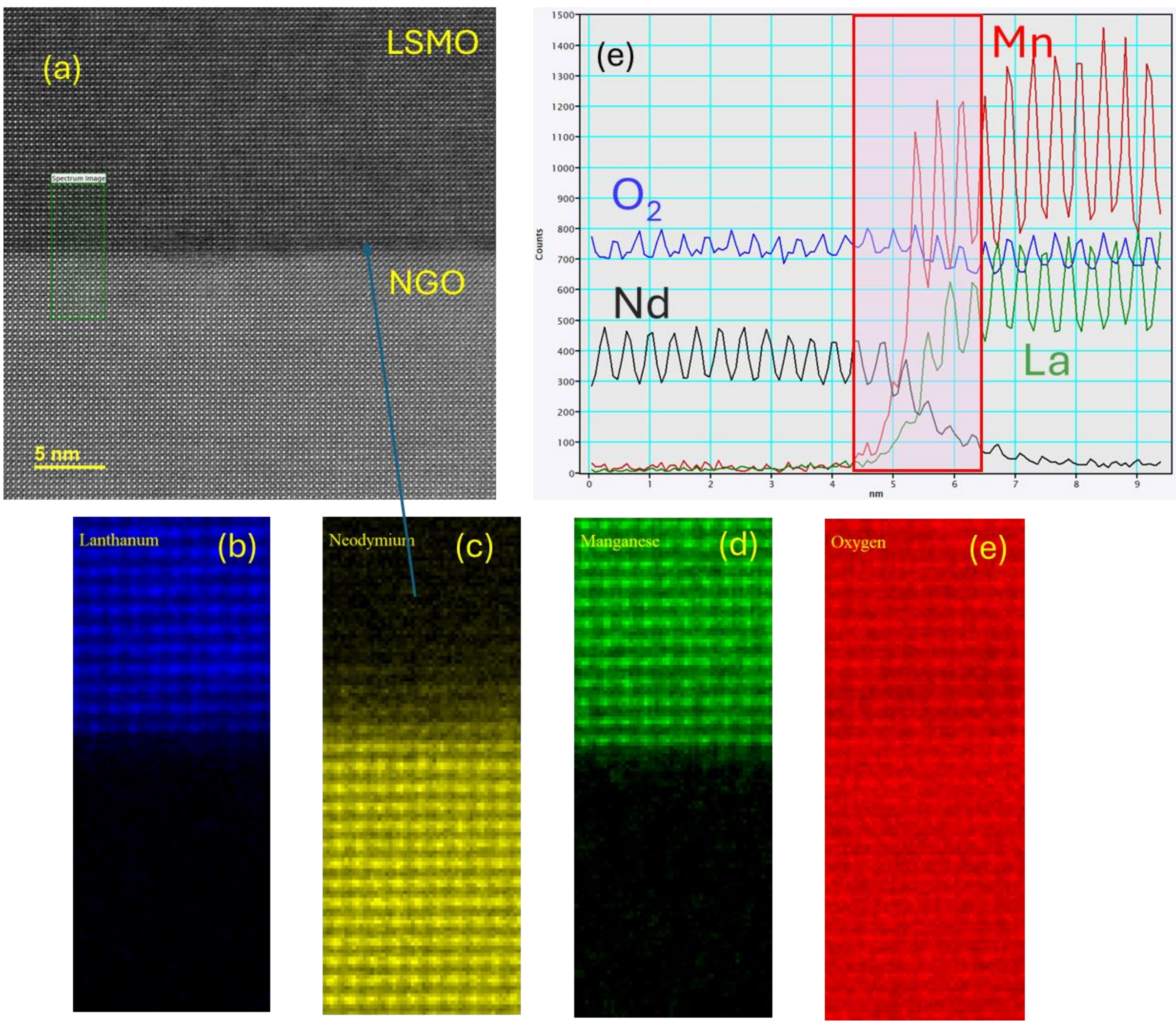

*Figure S4 (a) shows the zoomed in image of LSMO/NGO with the green square indicating the area where EELS measurement was done. (b)-(d) show the elemental contribution of La, Nd, Mn and O respectively. (c) shows*

The resistivity was measured as a function of temperature via the 4-probe method. This was performed in both the cooling and heating cycles to ensure no hysteretic effects were present in LSMO in our working range of temperature (Fig. S4).

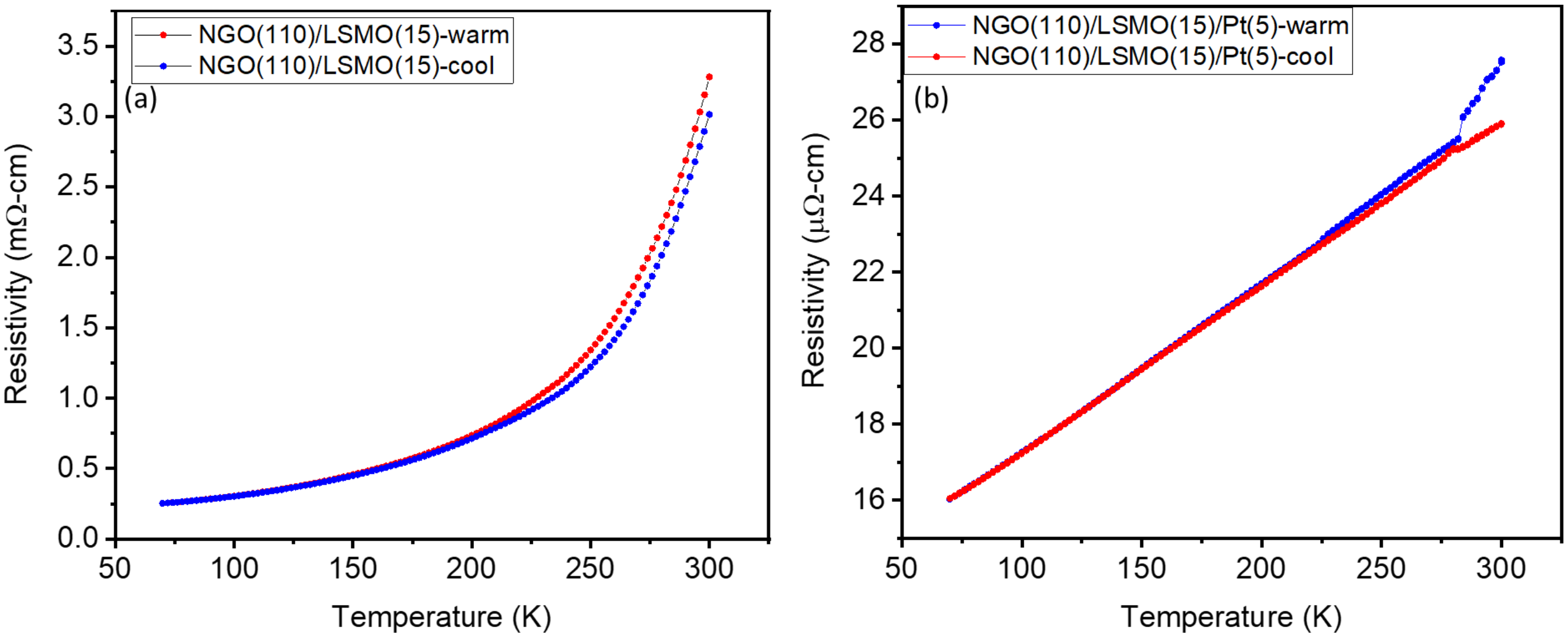

*Figure S4 . (a) and (b) shows the resistivity vs temperature measurements for S1 and S2. The slight offset may be due to movement of the probes during heating/cooling cycles. The values are similar to those reported in literature.*

The dispersion plots (Figure S5(a)-(b)) follow the usual square-root like trend for Hres vs frequency. As the temperature decreases, the saturation magnetization of LSMO increases resulting in lowered resonance frequencies and thus the backward shift of the plots. The lowering of the of the g-factor may indicate presence of orbital angular momentum contribution in the system Figure S5(c).

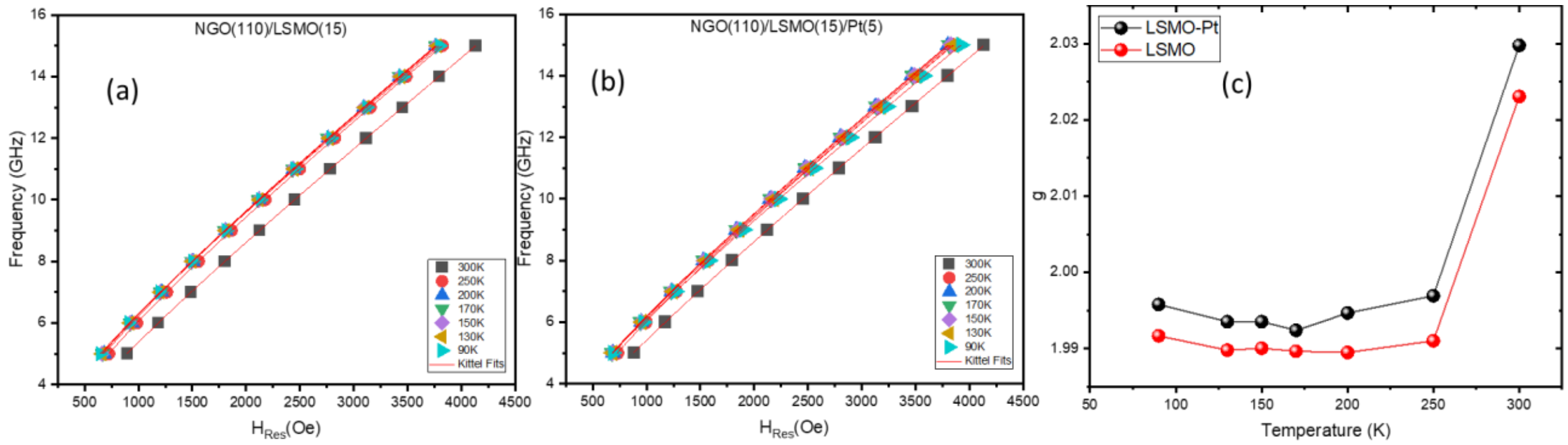

*Figure S5 Dispersion plots for S1 (a) and S2 (b), with the respective Kittel fittings to equation (2) of main manuscript. (c) shows the variation of the g-factor with temperature.*

The Kambersky's torque correlation models consider the relation of the electron relaxation time $\tau$ to the Gilbert damping parameter. When the spin waves propagate through the ferromagnet, there is a periodic deformation of the local Fermi surface. This causes redistribution of electrons amongst the varying fermi surface states, which is mediated by scattering processes whose lifetime is directly proportional to $\tau$ (ref 30 of main manuscript) In this process, when the role of spin orbit coupling is explicitly considered within the framework of band model of ferromagnetism, a dependence of the damping on the electrical conductivity emerges ( [1] [2]). This is the conductivity-like contribution ($\alpha_{intra}$) to the Gilbert damping, which depends on intra-band scattering processes. Additionally, the decay of spin waves generates electron-hole pairs with a finite lifetime due to the spin-orbit effects. This is a spin conserving process where a magnon is absorbed and flips the spin of the scattered electron. This inter-band scattering process is inversely proportional $\tau$ and scales with the resistivity. This leads to the resistivity-like term ($\alpha_{inter}$).

The overall expression for Kambersky's torque has the following form [3]

$$\alpha_{eff} = \alpha_{inter}\frac{\rho(T)}{\rho(300K)} + \alpha_{intra}\frac{\sigma(T)}{\sigma(300K)} \qquad (SE1)$$

From the fits to equation SE1 (Fig S6(a)) , $\alpha_{inter-S1(2)}$ for S1(S2), is found to be 0.00303 (0.00723) and $\alpha_{intra-S1(2)}$ was found to be $2.12\times 10^{-4}$($2.48\times 10^{-4}$) respectively. The values of $\alpha_{intra}$ is similar to those found in literature ( [4] [5]). Further, the inhomogeneous linewidth broadening is quite low for the thin films demonstrating good film quality.

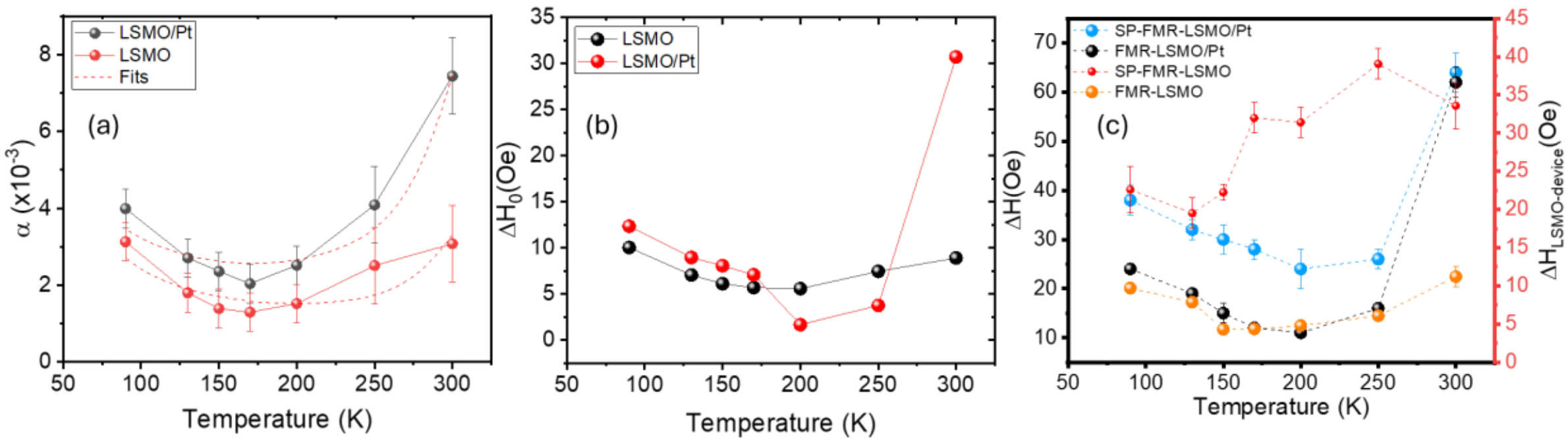

*Figure S6 (a) Damping vs temperature with fitting by equation SE1. (b) shows the inhomogeneous linewidth broadening. (c) shows the comparison of the linewidth for LSMO and LSMO/Pt samples measured via FMR and SP-FMR*

Temperature dependent magnetometry was performed using a vibrating sample magnetometer (VSM) by Quantum Design to confirm the ferromagnetic phase of the samples. Figure S7(a) shows the $M_s$ obtained from VSM vs calculated $M_{eff}$ from the dispersion plots. The decrease in the values from these two measurements may be due to the presence of an additional perpendicular anisotropy in the magnetically insulating layer. The increased in plane anisotropy field (Fig. S7 (c)) also reduces the contribution of the $M_{eff}$ causing it to saturate faster with reducing temperature. Such behavior of $H_K$ with temperature has been observed in LSMO [4] [5] and has been attributed to interaction between uncompensated spins at the interface of the metallic and PSMA LSMO layers.

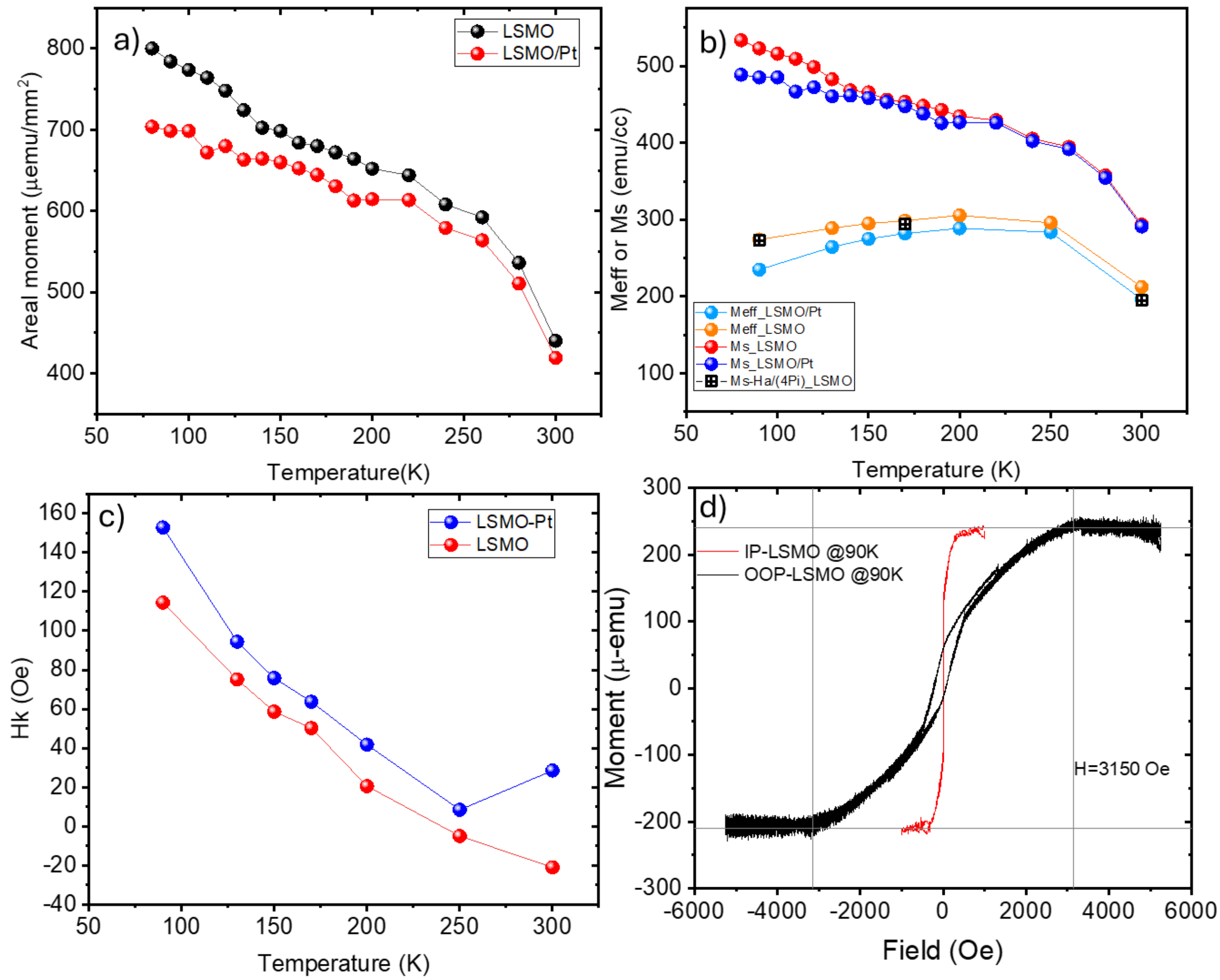

*Figure S7 (a)shows the areal moment of the LSMO and LSMO/Pt films> By comparing the ratio of the areal moments at room temperature, we obtain a dead layer of approx. 1nm for LSMO/Pt. (b) shows the variation of Meff obtained from FMR data in comparison to $M_s$ obtained from VSM measurements. The square boxed crosses indicate the values obtained of Meff from Ms-Ha/(4pi) from OOP MH measurements (c) shows the variation of the in-plane anisotropy field Hk obtained from FMR data, with respect to temperature (d) shows the in-plane and out of plane MH loops indicating an Ha of 3200 Oe at 90K.*

Figure S8 shows the comparison of Gilbert damping parameter and the linewidth (FWHM) for LSMO/Pt system with the popularly used Py/Pt system for oscillators. Both the linewidth and the damping are smaller in magnitude for the LSMO/Pt system. The damping is almost 8 times smaller in LSMO/Pt as opposed to Py/Pt at 160K. The linewidth is 5 times smaller for the above-mentioned same samples at 170K.

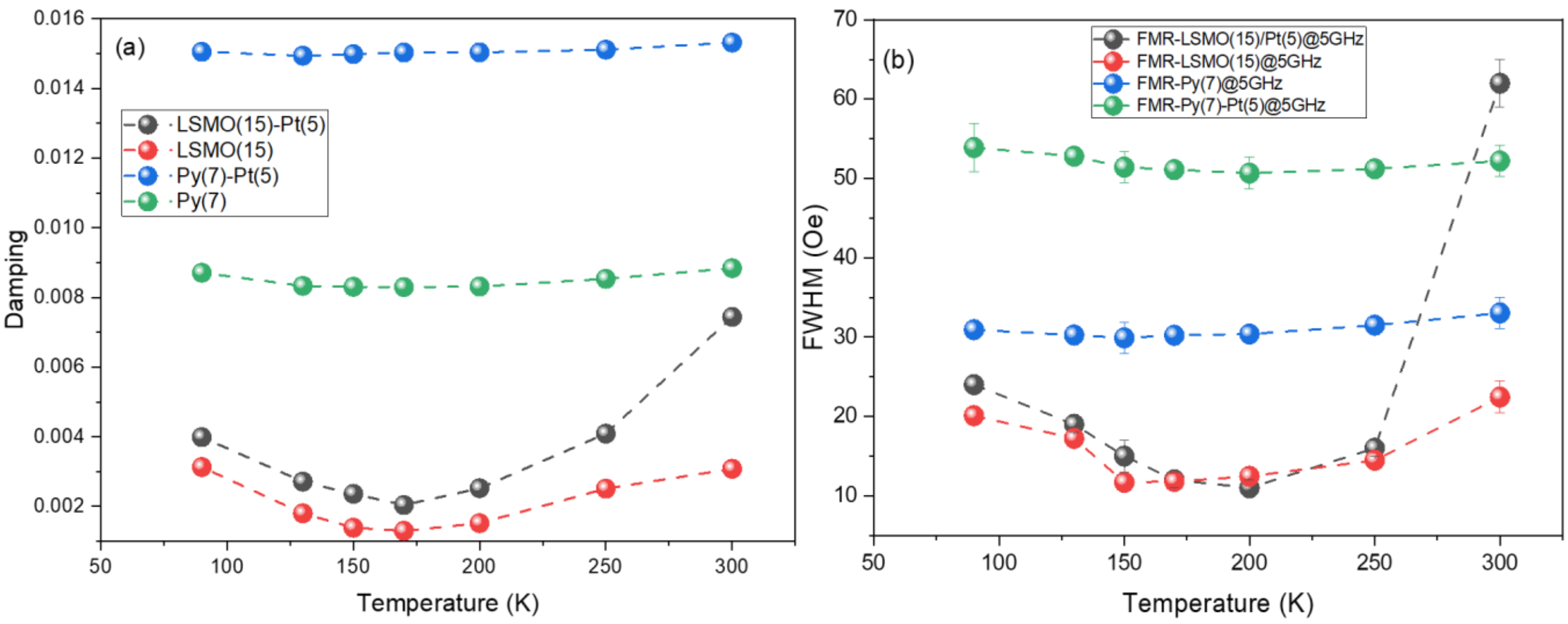

*Figure S8 (a) shows the variation of gilbert damping for Py(7),Py(7)/Pt(5), LSMO(15) and LSMO(15)/Pt(5) samples. (b) shows the variation of FWHM for the same samples obtained via FMR measurement at 5GHz.*